\documentclass[twocolumn,aps,prd,amsmath,eqsecnum,showpacs]{revtex4}

\usepackage{graphicx}

\begin{document}

\title
{Gravitational waves from a dust disk around a Schwarzschild black hole}
%
\author{Hajime Sotani}
\affiliation
{Research Institute for Science and Engineering, Waseda University,
Okubo 3-4-1, Shinjuku, Tokyo 169-8555, Japan}

\email{sotani@astro.auth.gr}
\altaffiliation
{\\Present address:
Department of Physics, Aristotle University of Thessaloniki,
Thessaloniki 54124, Greece.}
%
\author{Motoyuki Saijo}
\email{ms1@maths.soton.ac.uk}
\altaffiliation
{\\Present address:
School of Mathematics, University of Southampton, 
Southampton SO17 1BJ, United Kingdom.}
\affiliation{
Laboratoire de l'Univers et de ses Th\'eories,
Observatoire de Paris, F-92195 Meudon Cedex,  France
}
%
\received{6 July 2005}
\revised{22 May 2006}
\accepted{15 June 2006}
%
\begin{abstract}
We investigate gravitational waves from a dust disk around a
Schwarzschild black hole to focus on whether we can extract any of its
physical properties from a direct detection of gravitational waves.
We adopt a black hole perturbation approach in a time domain, which is
a satisfactory approximation to illustrate a dust disk in a
supermassive black hole.  We find that we can determine the radius of
the disk by using the power spectrum of gravitational waves and that
our method to extract the radius works for a disk of arbitrary density
distribution.  Therefore we believe a possibility exists for
determining the radius of the disk from a direct observation of
gravitational waves detected by the Laser Interferometer Space
Antenna.
\end{abstract}
\pacs{04.30.Db, 04.25.Nx, 04.70.-s}

\maketitle

\section{Introduction}
\label{sec:Intro}

Gravitational wave projects have been launched in many countries to
explore a new world in gravitational physics, in astrophysics, and in
cosmology.  Nowadays, several ground-based laser interferometric
gravitational wave detectors such as TAMA300, LIGO, GEO600, and VIRGO
have been seeking for a direct detection of gravitational
waves (See Ref. \cite{Barish05} for the current status of the
worldwide network of gravitational wave detectors).  The sensitive
frequency for the ground-based interferometers is within the range of
10 -- 1000 Hz.  The representative sources of gravitational waves are
the inspiral of binary systems composed of neutron stars and black
holes, spinning neutron stars, and low-mass x-ray binaries
\cite{CT02}.  There will be also a gravitational wave detector in 
space, Laser Interferometer Space Antenna (LISA).  The sensitive
frequency band in LISA is in the range of $10^{-4} - 10^{-1}$ Hz, and
one of the representative gravitational wave sources this detector is
studying is supermassive black holes whose masses are $M\sim 10^{6} -
10^{9} M_{\odot}$.  There is increasing evidence that a supermassive
black hole exists at the center of every galaxy \cite{Kormendy04}.

There is abundant observational evidence for dense gas in galactic
nuclei \citep[e.g.,][]{MF01,Kormendy04}.  For example, our galactic
nuclei contains a $4 \times 10^{6} M_{\odot}$ supermassive black hole
surrounded by an $\sim 10^{4} M_{\odot}$ molecular gas torus
\citep[e.g.,][]{JGGHMPST93}.  Also the disruption of a compact star by
a supermassive black hole spreads the gas in black hole spacetime
\cite{KHSHSP04, HGK04}.  From a theoretical point of view, the tidal
disruption of a star by a supermassive black hole has been studied in
Newtonian gravity \citep[e.g.,][]{Evans1989, CL83, CL85, Khokhlov1993,
  Marck1996} and in post-Newtonian gravity \cite{ALP00}, where the
black hole is regarded as a point mass.  There are also several
studies of tidal disruption in a Schwarzschild black hole
\cite{LMZD93, Mashhoon75} and in a Kerr black hole \cite{Fishbone73,
  FKNP94, Shibata96, DFKNP97}.  In all cases, the fragments of the
disrupted star are either swallowed by the supermassive black hole or
following a highly eccentric orbits.  There is also an indication,
based on an observation, that a disk can form around a supermassive
black hole \cite{CLG90}.

Black hole perturbation approach is one of the satisfactory tools to
illustrate gravitational waves from a compact star around a
supermassive black hole.  Many works have been published in
post-Newtonian expansion \cite{ST03}, in numerical analysis
\cite{NOK87} to compute gravitational waves from a test particle in a
Kerr black hole.  However all of the above works have been
investigated in a Fourier domain, which means that the orbital time of
the particle is infinite.  The basic equations of gravitational waves
are ordinary differential ones, but we have to treat ``infinite''
orbital time by restricting our orbit to a special one.  There is
another way of computing linearized Einstein equations: a time domain
approach, which directly solves the basic equations in partially
differential ones.  The main advantages of this time domain approach
are that we can impose an initial condition as an appropriate
distribution of the compact object and that we can treat an arbitrary 
orbit of the particle in general.  There are many studies in the field
of time domain computation of the head-on collision of two black holes
perturbatively \cite{LP97a,  LP97b, LP98}, of black hole perturbation
in Schwarzschild spacetime \cite{GPP94, MP02, Martel04, Poisson04,
  NFZP} and in Kerr spacetime \cite{KLPA97, LKP03, Khana04, AL05}, of
perturbation in spherical stars \cite{AK96, AAKS98, AAKLPR99, Ruoff01,
  RLP01}, of perturbation in nonrotating objects \cite{NDPF} and of
perturbation in slowly rotating stars \cite{RK01, RSK02}.  

Our purpose in this paper is to investigate whether we can extract any
physical property of the disk around a supermassive black hole from
gravitational waves.  Several studies have investigated extracting
physical properties of a tidally disrupted star in supermassive black
holes \cite{Vallisneri00, SN00, SN01, BEMSL04, KLPM04}.  When the disk
is formed from tidal disruption of a star by a supermassive black
hole, test particle approximation to illustrate the fragments is a
satisfactory tool since the pressure gradient is no longer dominant
\citep{KLPM04}.  \citet{SN00, SN01} investigated a tidally disrupted 
star falling into a black hole and found that we can extract the size
of the star from the energy spectrum of gravitational waves.  Although
they investigated gravitational radiation in a Fourier domain,
inappropriate description of the pre-disruption of the star, their
Fourier domain approach does almost illustrate the picture of
post-disruption appropriately since most gravitational waves are
radiated at a quasi-normal ringing phase.  Here we mainly focus on
gravitational waves from a disrupted star around a supermassive black
hole in a time domain to learn whether we can extract any physical
property of a dust disk.  We have to deal with two types of test
particle orbit in our dust disk calculation due to the angular
momentum depletion.  Particles both plunging into and orbiting a
Schwarzschild black in the same calculation.  To perform the above
situation from the computational point of view, time domain approach
is quite easy to handle rather than Fourier domain one.

This paper is organized as follows.  In Sec.~\ref{sec:II}, we briefly
describe the basic equations of gravitational waves from a test
particle.  In Sec.~\ref{sec:III}, we demonstrate gravitational waves
from a test particle in our developed code, comparing with well-known
results in this field.  In Sec.~\ref{sec:IV}, we show gravitational
waves from a dust disk around a Schwarzschild black hole.  We
summarize our paper in Sec.~\ref{sec:V}.  Throughout this paper, we
use geometrized units $G=c=1$, where $G$ and $c$ denote the
gravitational constant and the speed of light, respectively, and the 
metric sign as $(-,+,+,+)$.

\section{Black hole perturbation approach in Schwarzschild spacetime}
\label{sec:II}

Here we describe the basic equations of gravitational waves in order
to compute gravitational waves from a test particle around a
Schwarzschild black hole.

First, we describe the motion of a test particle with a rest mass
$\mu$ around a Schwarzschild black hole with a gravitational mass $M$.
In this paper, we only consider the motion of a test particle in the
equatorial plane.  The motion is specified by giving the two parameters
$\tilde{E}_{\rm p}$ and $\tilde{L}_{\rm p}$ in units of $\mu$ as
\begin{align}
 \tilde{E}_{\rm p} = E_{\rm p}/\mu,\ \ \tilde{L}_{\rm p} = L_{\rm p}/\mu,
\end{align}
where $E_{\rm p}$ and $L_{\rm p}$ are the energy and the total angular
momentum of a test particle observed at infinity, respectively.  With
these parameters, the equation of motion of a test particle is
described as 
\begin{align}
\frac{d^2 r_*}{dt^2} &= 
  \frac{1}{\tilde{E}_{\rm p}^{\ 2}}\left(1-\frac{2M}{r}\right)
  \left[
     -\frac{M}{r^2}+\frac{\tilde{L}_{\rm p}^{\ 2}}{r^3}
     \left(1-\frac{3M}{r}\right)\right],
\label{eqn:eomr} \\
\frac{d\phi}{dt} &= 
  \frac{1}{r^2}
  \left(1-\frac{2M}{r}\right)
  \frac{\tilde{L}_{\rm p}}{\tilde{E}_{\rm p}},
\end{align}
where $r_*=r+2M\ln (r/2M-1)$ is the tortoise coordinate.  We also
introduce the effective potential $\tilde{V}_{p}$ by integrating the
radial motion of a test particle (Eq.~[\ref{eqn:eomr}]) as
\begin{align}
\left(\frac{dr_*}{dt}\right)^2 &= 
1 - \left(\frac{\tilde{V}_{\rm p}(r)}{\tilde{E}_p}\right)^2,
\label{dvdt2} \\
\tilde{V}_p(r) &= 
  \sqrt{\left(1-\frac{2M}{r}\right)
  \left(1+\frac{\tilde{L}_{\rm p}^{\ 2}}{r^2}\right)}.
\label{potentialV}
\end{align}
The potential $\tilde{V}_{\rm p}$ has a local minimum when
$\tilde{L}_{\rm p} \geq 2 \sqrt{3} M$ at the radius
\begin{equation}
 r = 
\frac{\tilde{L}_{\rm p}}{2M}
\left(\tilde{L}_{\rm p}+\sqrt{\tilde{L}_{\rm p}^{\ 2}-12M^2}\right),
\end{equation}
where the particle has a stable circular orbit.

Next, we describe the basic equations of gravitational waves.  These
equations are obtained by linearizing the Einstein equations in
Schwarzschild spacetime called Regge-Wheeler equations for odd parity
\cite{RW57} and Zerilli equations for even parity \cite{Zerilli70}.
Note that we can only treat odd and even parity separately in the
nature of spherical symmetric background spacetime.

The radial wave function of gravitational waves obeys
Regge-Wheeler-Zerilli equations as
\begin{eqnarray}
&&\left[
  -\frac{\partial^2}{\partial t^2}+\frac{\partial^2}{\partial r_*^2}-
  V_l(r)^{\rm (odd/even)}
\right] 
\Psi_{lm}^{\rm (odd/even)}(t,r)
\nonumber \\ 
&&~~~~~
= S_{lm}^{\rm (odd/even)}(t,r), 
\label{RWZ}
\end{eqnarray}
where $\Psi_{lm}^{\rm (odd/even)}$, $S_{lm}^{\rm (odd/even)}$, $V_{l}^{\rm
  (odd/even)}$ is the radial wave function, source term, and effective 
potential of gravitational waves, respectively.  The effective
potentials for odd and even parities are given by
\begin{align}
V_{l}^{\rm (odd)} &= 
  \frac{2f}{r^2}\left(\lambda+1-\frac{3M}{r}\right),  \\
V_{l}^{\rm (even)} &= 
  \frac{2f\left(\lambda^2 (\lambda+1)r^3+3M\lambda^2 r^2+
            9M^2\lambda r+9M^3\right)}{r^5\Lambda^2},
\end{align}
where $f\equiv 1-2M/r$, $\Lambda\equiv\lambda+3M/r$, and $\lambda
\equiv (l+2)(l-1)/2$.  The source terms are constructed from the
energy momentum tensor of a test particle as
\begin{align}
 S_{lm}^{\rm (odd)} &= 
  -\frac{8\pi rf}{\sqrt{2\lambda(\lambda+1)}}
            \Bigg[\frac{d}{dr}\left(fD_{lm}\right)
         - \frac{\sqrt{2\lambda}f}{r}Q_{lm}\Bigg],
\\
 S_{lm}^{\rm (even)} &= 
  \frac{8\pi}{(\lambda+1)\Lambda}
           \Bigg[r^2f\left(f^2\frac{\partial}{\partial r}
    (f^{-2}A_{lm}^{(0)})-\frac{\partial}{\partial r}(f^2A_{lm})\right)
\nonumber \\
 & + rf^2(\Lambda-f)A_{lm}
    +\sqrt{2}rf^2G_{lm} 
\nonumber \\
 & -\frac{1}{r\Lambda} 
    [
      \lambda(\lambda-1)r^2+(4\lambda-9)Mr+15M^2
    ]
    A_{lm}^{(0)}\Bigg]
\nonumber \\
 & +\frac{16\pi rf^2}{\Lambda\sqrt{\lambda+1}}B_{lm}-
     \frac{16\pi rf}{\sqrt{2\lambda(\lambda+1)}}F_{lm},
\end{align}
where
\begin{align}
 A^{(0)}_{lm} &
=\mu\gamma f^2\frac{\delta(r-\hat{R})}{r^2}\,\bar{Y}_{lm}(\hat{\Omega}), \\
 A_{lm}       &
=\mu\gamma\left(\frac{d\hat{R}}{dt}\right)^2
 \frac{\delta(r-\hat{R})}{(r-2M)^2}\,
 \bar{Y}_{lm}(\hat{\Omega}), \\
 B_{lm}       &
=\mu\gamma\sqrt{\frac{1}{\lambda+1}}\frac{d\hat{R}}{dt}
 \frac{\delta(r-\hat{R})}{r-2M}\,
 \frac{d\bar{Y}_{lm}(\hat{\Omega})}{dt}, \\
 D_{lm}       &
=\frac{i\mu\gamma\delta(r-\hat{R})}{2\sqrt{2\lambda(\lambda+1)}}
 \left(\frac{d\hat{\Phi}}{dt}\right)^2 \bar{X}_{lm}(\hat{\Omega}), \\
 F_{lm}       &
=-\frac{\mu\gamma\delta(r-\hat{R})}{2\sqrt{2\lambda(\lambda+1)}}
 \left(\frac{d\hat{\Phi}}{dt}\right)^2 \bar{W}_{lm}(\hat{\Omega}), \\
 G_{lm}       &
=\frac{\mu\gamma}{\sqrt{2}}\delta(r-\hat{R})
 \left(\frac{d\hat{\Phi}}{dt}\right)^2 \bar{Y}_{lm}(\hat{\Omega}), \\
 Q_{lm}       &
=i\mu\gamma\sqrt{\frac{1}{\lambda(\lambda+1)}}\frac{\delta(r-\hat{R})}{r-2M}
 \frac{d\hat{R}}{dt} \frac{d\hat{\Phi}}{dt}
 \frac{\partial Y_{lm}(\hat{\Omega})}{\partial\theta}.
\end{align}
Note that $A^{(0)}_{lm}$, $A_{lm}$, $B_{lm}$, $D_{lm}$, $F_{lm}$,
$G_{lm}$, and $Q_{lm}$ are the coefficients expanded in tensor
harmonics \cite{Zerilli70}.  Note also that $\gamma\equiv dt/d\tau =
\tilde{E}_{\rm p} / (1-2M/r)$, where $\tau$ is a proper time of the
particle.  $X^i(t) = \left( \hat{R}(t), \hat{\Omega}(t) \right)$ is
the location of a particle, where $\hat{\Omega} = \left(\pi/2,
\hat{\Phi}(t)\right)$.  Also a quantity $\bar{A}$ represents the
complex conjugate of a quantity $A$.  The functions $X_{lm}$ and
$W_{lm}$ are the tensorial part of tensor harmonics defined by
\begin{align}
 X_{lm} &= 2\partial_{\phi}(\partial_{\theta}-\cot\theta)Y_{lm}, \\
 W_{lm} &= \left(\partial_{\theta}^2-\cot\theta\partial_{\theta}
         -\frac{\partial_{\phi}^2}{\sin^2\theta}\right)Y_{lm}.
\end{align}

Finally, we mention the relationship between the radial wave function
and the perturbed metric in the wave zone.  It is written as
[Eq.~(3.1) of Ref.~\cite{Martel04}]
\begin{eqnarray}
h_{+} - i h_{\times} &=& \frac{1}{2r} \sum_{l,m} 
\sqrt{\frac{(l+2)!}{(l-2)!}}
{}_{-2}Y_{lm}(\theta, \varphi)
\nonumber \\
&& \times
\left(
  \Psi_{lm}^{\rm (even)} - 2 i \int^{t} dt' \Psi_{lm}^{\rm (odd)}
\right)
\label{eqn:h-Psi}
,
\end{eqnarray}
where spin-weighted spherical harmonics ${}_{-2}Y_{lm}(\theta,
\varphi)$ is defined by 
\begin{equation}
{}_{-2}Y_{lm}(\theta,\varphi) = \frac{1}{2\sqrt{\lambda (\lambda + 1)}}
\left(
  W_{lm} - \frac{i}{\sin \theta} X_{lm}
\right)
.
\end{equation}
The energy flux ($dE_{\rm gw}/dt$) of gravitational waves at the wave
zone is also given by [Eq.~(B.26) of Ref.~\cite{Martel04}]
\begin{align}
\frac{dE_{\rm gw}}{dt} =\frac{1}{64\pi}\sum_{l,m}\frac{(l+2)!}{(l-2)!}
\left[
  \left| \frac{\partial}{\partial t} {\Psi}_{lm}^{\rm (even)}\right|^2+
  4\left|\Psi_{lm}^{\rm (odd)}\right|^2
\right].
\label{energy-flux}
\end{align}

\section{Gravitational waves from a test particle in Schwarzschild spacetime}
\label{sec:III}
Here we compute gravitational waves from a test particle in
Schwarzschild spacetime.  We use a time domain approach to solve
Eq.~(\ref{RWZ}).  In order to test our newly developed code, we
compare our results in two cases with well-known semi-analytical
results.  One is gravitational waves from a particle in a circular
orbit around a Schwarzschild black hole, which has been studied up to
the 5.5 post-Newtonian order \cite{TTS96}.  The other is gravitational
waves from a particle with a radial in-fall into a Schwarzschild black
hole in a Fourier domain \cite{DRPP71}.

We set the initial condition of the radial wave function as 
\begin{align}
\left. \Psi_{lm}^{\rm (odd/even)} \right|_{t=0} = 0,
\ \ 
\frac{\partial\Psi_{lm}^{\rm (odd/even)}}{\partial t}\bigg|_{t=0}=0,
\label{eqn:initial}
\end{align}
with the boundary condition that there is no incoming wave to our
system.  Namely, we impose the outgoing wave boundary condition for
the radial wave function at horizon and at a spatial 
infinity as 
\begin{eqnarray}
 \left(\frac{\partial}{\partial t} + \frac{\partial}{\partial r_*}\right)
    \Psi^{\rm{(odd/even)}}_{lm} = 0
    \ \ &{\mbox{at}}&\ r_* = r_*^{\rm in},
\label{eqn:bd1}
\\
 \left(\frac{\partial}{\partial t} - \frac{\partial}{\partial r_*}\right)
    \Psi^{\rm{(odd/even)}}_{lm} = 0
    \ \ &{\mbox{at}}&\ r_* = r_*^{\rm out},
\label{eqn:bd2}
\end{eqnarray}
where $r_{*}^{\rm in}$ and $r_{*}^{\rm out}$ are the inner and outer edge
of the computational grid.  In practice  Eqs.~(\ref{eqn:bd1}) and
(\ref{eqn:bd2}) are written as 
\begin{align}
\Psi^{\rm{(odd/even)}}_{lm}(t + \Delta t, r^{\rm in}_{*}) =&
  \left(1 - \frac{\Delta t}{\Delta r_*}\right) 
    \Psi^{\rm{(odd/even)}}_{lm}(t, r^{\rm in}_{*}) \nonumber \\
  &+ \frac{\Delta t}{\Delta r_*} 
    \Psi^{\rm{(odd/even)}}_{lm}(t, r^{\rm in}_{*}+\Delta r_{*}), 
\\
\Psi^{\rm{(odd/even)}}_{lm}(t + \Delta t, r^{\rm out}_{*}) =&
  \left(1 - \frac{\Delta t}{\Delta r_*}\right) 
    \Psi^{\rm{(odd/even)}}_{lm}(t, r^{\rm out}_{*}) \nonumber \\
  &+ \frac{\Delta t}{\Delta r_*} 
    \Psi^{\rm{(odd/even)}}_{lm}(t, r^{\rm out}_{*}-\Delta r_*)
, 
\end{align}
where $\Delta t$ and $\Delta r_*$ are the step-size for time and space
respectively.  Note that we apply the first order finite differencing
scheme in both space and time to the boundary conditions.

Although the initial condition we choose (Eq.~[\ref{eqn:initial}]) is
somewhat artificial, the particle motion is unaffected by
gravitational radiation in our test particle approximation, and hence,
after a certain evolution time, the radial wave function should
converge into an appropriate solution.  Therefore, we simply ignore
the early stage of the evolution throughout this paper.

We should carefully deal with the delta function ${\delta(r-\hat{R})}$,
which appears in the source term [Eq.~(\ref{RWZ})].  One treatment is
to approximate the delta function to a step function that only covers
one grid cell in the null coordinates \cite{LP97b, MP02}.  The other
is to adopt, as we do here, an approximate function to describe a
delta function, which appears in the source term, as 
\cite{RLP01}
\begin{equation}
\delta (r - \hat{R}) = \frac{1}{\sqrt{\pi} \sigma} 
\exp \left[ -\frac{(r-\hat{R})^{2}}{\sigma^{2}} \right]
,
\label{eqn:gaussian}
\end{equation}
where $\sigma$ is a standard deviation parameter.  Note that
Eq.~(\ref{eqn:gaussian}) becomes exactly a delta function in the limit
of $\sigma \rightarrow 0$.

We solve Eq.~(\ref{RWZ}) in second order accuracy in space and time
with the scheme based on two times iterated Crank-Nicholson method
\cite{NR, Teukolsky00}.  We set the computational grid as $-1000M \le
r_* \le R_{\rm obs}+2000M$, where $R_{\rm obs}$ is the location of the
observer in the equatorial plane.  We also set the step-size for space
and time as  $\Delta r_* = 1.0 M$ and $\Delta t = 0.5 \times \Delta
r_{*}$, which satisfies the Courant condition.  Although we vary the
location of the inner boundary $r_*\le-1000M$, the difference of
gravitational wave amplitude is within the round-off error.

We set the standard deviation parameter $\sigma$ as $\sigma = \Delta
r_*$.  When $\sigma$ is less than $\Delta r_*$, it is not certain to
specify the location of a test particle due to the less spatial grid
resolution, and therefore we discard such region.  More concrete
examinations of the dependence of $\sigma$ on the circular and free
fall orbits are discussed in the next two subsections.

\subsection{Circular orbits}
\label{sec:III-1}

\begin{center}
\begin{figure}[htbp]
\includegraphics[width=0.45\textwidth]{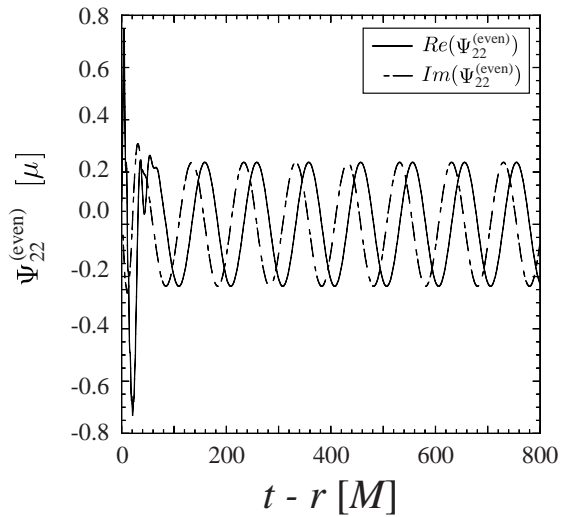}
\caption{
Radial wave function of a test particle in a circular orbit at
$R_c=10M$.  We only show the $l=m=2$ mode, which is the dominant mode
according to the observer in the equatorial plane at $R_{\rm
  obs}=2000M$.  We set the standard deviation parameter as  $\sigma=0.1M$.
Solid and dash-dotted lines represent real and imaginary parts of the
radial wave function, respectively.
}
\label{fig-wave-circular1}
\end{figure}
\end{center}

We compute the radial wave function from a test particle in a circular 
orbit at  $R_{\rm c}=10M$ in Fig.~\ref{fig-wave-circular1}, where
$R_{\rm c}$ is the radius of the orbit.  We set the standard deviation
parameter as $\sigma=0.1M$ and set the location of the observer at
$R_{\rm obs}=2000M$.  We only show the $l=m=2$ mode, which is the
dominant one for the observer in the equatorial plane.  We find a
burst in the early stage of the evolution, which comes from the
inappropriate initial condition we imposed (Eq.~[\ref{eqn:initial}]).
Since the source term of Eq.~(\ref{RWZ}) is unaffected by our initial
condition, we simply ignore the unphysical burst in the wave function.

\begin{center}
\begin{figure}[htbp]
\includegraphics[width=0.45\textwidth]{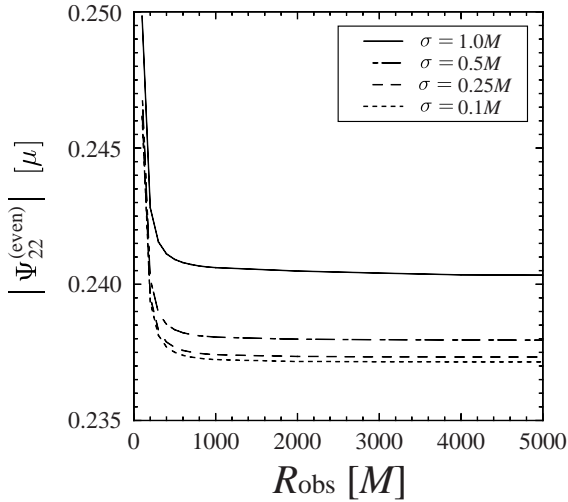}
\caption{
The amplitude of the radial wave function as a function of the
location of the observer $R_{\rm obs}$.  We only show the $l=m=2$ mode.
Solid, dash-dotted, dashed, dotted lines denote the standard deviation
parameters $\sigma = 1.0 M$, $0.5 M$, $0.25M$, $0.1M$, respectively.
}
\label{dep-circular-observer}
\end{figure}
\end{center}

We vary the location of the observer $R_{\rm obs}$ to investigate the
radius which is numerically regarded as infinity in
Fig.~\ref{dep-circular-observer}.  We only compute $l=m=2$ mode, and 
vary the standard deviation parameter $\sigma$.  It is clearly shown in 
Fig.~\ref{dep-circular-observer} that the amplitude becomes almost
constant when $R_{\rm obs} \ge 2000M$, irrespective of the standard 
deviation parameter $\sigma$.  Hereafter, we set the location of the 
observer at $R_{\rm obs} \ge 2000M$.

We compare our numerical results with the ones of 5.5 post-Newtonian
expansion \cite{TTS96} at two different radii, i.e., $R_c = 7 M$
and $10 M$.  The energy flux of gravitational waves $(dE_{\rm
  gw}/dt)^{\rm (PN)}$ from the analytical calculation (Eq.~[3.1] of
Ref.~\cite{TTS96}) is that
\begin{equation}
\left( \frac{dE_{\rm gw}}{dt} \right)^{\rm (PN)}
=
\left\{ 
\begin{array}{ll}
3.92 \times 10^{-5} ( \mu / M )^2 & 
\mbox{for}\  R_{\rm c} = 7 M.
\\
6.13 \times 10^{-5} ( \mu / M )^2 & 
\mbox{for}\  R_{\rm c} = 10 M.
\end{array}
\right.
\label{eqn:PNexpand}
\end{equation}
Our energy flux for four different standard deviation parameters
$\sigma$ at $R_{\rm obs}=2000M$ is summarized in
Table~\ref{tab-circular}.  We sum up the index of the spherical
harmonics $l$ from $2$ to $6$.  From Table~\ref{tab-circular}, we find
that the deviation from the results of 5.5 post-Newtonian
expansion becomes small when the standard deviation parameter $\sigma$
becomes small.  It is certainly true that this difference disappears
within the limit of $\sigma \rightarrow 0$.  We therefore conclude
that for a circular orbit our code has good accuracy for $\sigma
\lesssim 0.5 M$ within the relative error of few percent in the energy
flux.
\begin{table}[htbp]
      \begin{center}
          \leavevmode
          \caption{
Comparison of our numerical results with the results of 
5.5 post-Newtonian expansion (Eq.~[\ref{eqn:PNexpand}]).
          }
          \begin{tabular}{c c c c c c}
\hline\hline
  & $R_{\rm c}$\footnotemark[1] & $\sigma$\footnotemark[2] & 
    $dE_{\rm gw}/dt$ $[\mu^2/M^2]$\footnotemark[3] & 
    relative error [\%] & \\
\hline
  &  $7 M$ & $1.00M$  & $4.15\times 10^{-5}$  & 5.87 & \\
  &  $7 M$ & $0.50M$  & $4.04\times 10^{-5}$  & 3.06 & \\
  &  $7 M$ & $0.25M$  & $4.01\times 10^{-5}$  & 2.30 & \\
  &  $7 M$ & $0.10M$  & $4.00\times 10^{-5}$  & 2.04 & \\
\hline
  & $10 M$ & $1.00M$  & $6.35\times 10^{-5}$  & 3.59 & \\
  & $10 M$ & $0.50M$  & $6.20\times 10^{-5}$  & 1.14 & \\
  & $10 M$ & $0.25M$  & $6.16\times 10^{-5}$  & 0.489 & \\
  & $10 M$ & $0.10M$  & $6.15\times 10^{-5}$  & 0.326 & \\
\hline\hline
          \end{tabular}
          \label{tab-circular}
\footnotetext[1]{$R_{\rm c}$: Radius of the orbit}
\footnotetext[2]{$\sigma$: Standard deviation parameter}
\footnotetext[3]{$dE_{\rm gw}/dt$: Energy flux of gravitational waves}
\end{center}
\end{table}

We also mention the convergence order of our code.  Since we use
finite differencing scheme and approximate the delta function, there
are two parameters $\sigma$ and $\Delta r_*$ for investigating the
convergence order.  

First we fix $\sigma$ as $0.5 M$ and varies $\Delta r_*$ between
$0.1M$ and $0.5M$.  The energy flux ($dE_{\rm gw}/dt$) at $R_{\rm c} =
10M$ are 
$6.17\times 10^{-5} [\mu^2/M^2]$ for $\Delta r_*=0.1M$, 
$6.17\times 10^{-5} [\mu^2/M^2]$ for $\Delta r_*=0.2M$
(relative error from the energy flux of $\Delta r_* = 0.1M$ is
$0.07$\%),
$6.19\times 10^{-5} [\mu^2/M^2]$ for $\Delta r_*=0.4M$
(relative error $0.35$\%), and 
$6.20\times 10^{-5} [\mu^2/M^2]$ for $\Delta r_*=0.5M$ 
(relative error $0.56$\%). 
Therefore our code roughly has second order convergence in space,
which is consistent with our second order finite differencing scheme. 
Next we fix the relationship $\sigma = \Delta r_*$ and varies $\sigma$
between $0.1M$ and $1.0 M$.  Roughly speaking our code has first order
convergence in $\sigma$, since the relative error in
Table~\ref{tab-circular} decreases one half as $\sigma$ becomes one
half.

\subsection{Free fall orbits}
\label{sec:III-2}

\begin{center}
\begin{figure}[htbp]
\includegraphics[width=0.50\textwidth]{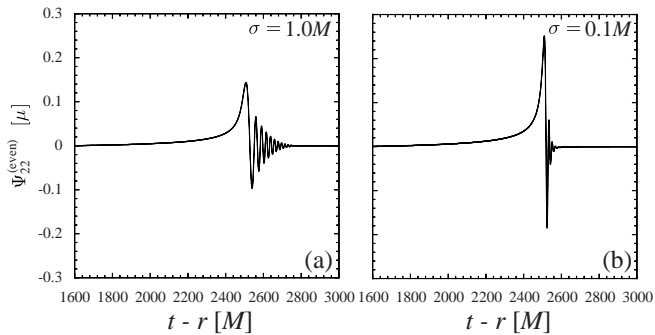}
\caption{
Radial wave function of a test particle falling radially into a 
Schwarzschild black hole at $R_{\rm obs}=7000M$.  We only show the
$l=m=2$ mode.  The standard deviation parameters $\sigma$ are (a)
$1.0M$ and (b) $0.1M$, respectively.  The parameters of oscillation
frequency and the damping rate of the ringing tail are ($M\omega$,
$M/\tau$) = (a) ($0.199$, $0.015$), (b) ($0.314$, $0.063$),
respectively.
}
\label{fig-wave-freefall1}
\end{figure}
\end{center}

Next we compute gravitational waves from a test particle falling into
a Schwarzschild black hole.  We show the radial wave function of
$l=m=2$ mode at $R_{\rm obs}=7000M$ in Fig.~\ref{fig-wave-freefall1}.
Instead of following the geodesics of a particle from infinity, we put
an in-fall velocity at $r = r_{\rm int}$ of $v \equiv
dr_*/dt=-\sqrt{2M/r_{\rm int}}$ derived from Eqs.~(\ref{dvdt2}) and 
(\ref{potentialV}).   We also vary the initial radius of the test
particle $r_{\rm int}$ in the range of $200 M \le r_{\rm int} \le 3000
M$, and find that the total amount of radiated energy is in the
relative error range of 0.001\%.  The radiated energy is determined by
the integration of Eq.~(\ref{energy-flux}) in the range of $1600 M
\lesssim t-r \lesssim 3000 M$ in Fig.~\ref{fig-wave-freefall1}.  We
set the spatial grid size $\Delta r_* = \sigma$, as we mentioned
before.  We show our convergence test of the ringing tail, varying
$\Delta r_*$ in Table \ref{tab-omega}.  We find that our computation
has a convergence when we take $\sigma < 0.25 M$.  The resonance
frequency agrees with the early results of \citet{SN82} ($M \omega
\approx 0.30$), which corresponds to the frequency of the quasinormal
mode of a Schwarzschild black hole.
\begin{table}[htbp]
      \begin{center}
          \leavevmode
          \caption{
Resonance frequencies of the quasi-normal ringing
for some values of $\Delta r_*$ and $\sigma$.
          }
          \begin{tabular}{ccccc}
            \hline\hline
& $\Delta r_*$
& $\sigma$
& $M \omega$ & \\
\hline
              & $0.10M$  & $0.10M$  & $0.31$ &  \\
              & $0.25M$  & $0.25M$  & $0.30$ &  \\
              & $0.50M$  & $0.50M$  & $0.29$ &  \\
              & $1.00M$  & $1.00M$  & $0.20$ &  \\
            \hline\hline
          \end{tabular}
          \label{tab-omega}
      \end{center}
\end{table}
\begin{table}[htbp]
      \begin{center}
          \leavevmode
          \caption{
Comparison of our numerical results with the results of a free
fall orbit ($l=2$ mode) \cite{DRPP71}.
          }
          \begin{tabular}{cccccc}
            \hline\hline
        & $R_{\rm obs}$ & $\sigma$ & $E_{\rm gw}$ $[\mu^2/M]$ & 
          relative error [\%] & \\
\hline
        & $5000 M$ & $1.00M$ & $4.94 \times 10^{-3}$  & 46.3 & \\
        & $5000 M$ & $0.50M$ & $8.38 \times 10^{-3}$  & 8.91 & \\
        & $5000 M$ & $0.25M$ & $8.78 \times 10^{-3}$  & 4.57 & \\
        & $5000 M$ & $0.10M$ & $8.96 \times 10^{-3}$  & 2.61 & \\
\hline
        & $7000 M$ & $1.00M$ & $4.17 \times 10^{-3}$  & 54.7 & \\
        & $7000 M$ & $0.50M$ & $7.94 \times 10^{-3}$  & 13.7 & \\
        & $7000 M$ & $0.25M$ & $8.71 \times 10^{-3}$  & 5.33 & \\
        & $7000 M$ & $0.10M$ & $8.95 \times 10^{-3}$  & 2.72 & \\

            \hline\hline
          \end{tabular}
          \label{tab-freefall}
      \end{center}
\end{table}

We compare our result with well-known one calculated in a frequency
domain \cite{DRPP71}.  The radiated energy $E_{\rm gw}$ for the $l=2$
mode is $E_{\rm gw} = 9.2 \times 10^{-3} (\mu^2/M)$ in
Ref.~\cite{DRPP71}.  Our results are summarized in Table
\ref{tab-freefall}.  We have the same behavior as in the case of a
circular orbit that the error is small as the standard deviation parameter
becomes small.  However, the relative error is larger than in the case
of a circular orbit because a particle moves close to the black hole
horizon, where the gravitational field is strong.

\begin{center}
\begin{figure}[htbp]
\includegraphics[width=0.45\textwidth]{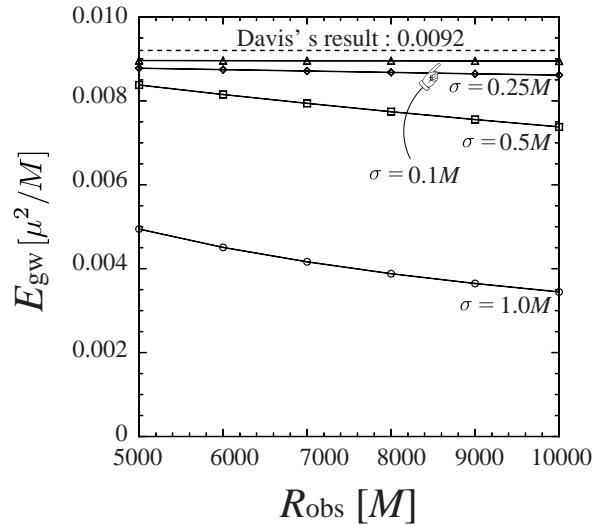}
\caption{
Radiated energy of gravitational waves from a radially in-falling
particle into a Schwarzschild black hole for the $l=2$ mode.  Circle,
square, diamond, and triangle denote $\sigma=1.0M$, $0.5M$, $0.25M$,
and $0.1M$, respectively.  Dotted line represents the numerical
result of Ref. \cite{DRPP71}.
}
\label{dep-freefall-observer}
\end{figure}
\end{center}

We also investigate the convergence order of our code in the free fall
orbit.  First we fix $\sigma$ as $0.5 M$ and varies $\Delta r_*$
between $0.1M$ and $0.5M$.  The radiated energy $E_{\rm gw}$ at
$R_{\rm obs} = 7000M$ are  
$9.65 \times 10^{-3} [\mu^2/M]$ for $\Delta r_*=0.1M$, 
$9.54 \times 10^{-3} [\mu^2/M]$ for $\Delta r_*=0.2M$
(relative error from the radiated energy of $\Delta r_* = 0.1M$ is
$1.17$\%), 
$8.70 \times 10^{-3} [\mu^2/M]$ for $\Delta r_*=0.4M$
(relative error $9.92$\%), and 
$7.94 \times 10^{-3} [\mu^2/M]$ for $\Delta r_*=0.5M$
(relative error $17.7$\%).
Therefore our code has second order convergence in space.  Next we fix
the relationship $\sigma = \Delta r_*$ and varies $\sigma$ between
$0.1M$ and $1 M$.  Roughly speaking our code has a first order
convergence in $\sigma$, since the relative error in
Table~\ref{tab-freefall} decreases  one half as $\sigma$ becomes
one half.  Note that the case $\sigma = 1.0 M$ may be out of the
convergence region.

We also set the observer quite far from the source in order to
distinguish easily between the unphysical burst and the physical
waveform.  In fact, we integrate to compute the radiated energy in the
range of $1600 M <  t-r_* < 3000 M$.  Note that the radiated energy is
irrespective of the variation of the range of integration.  In
Fig.~\ref{dep-freefall-observer}, we also show the dependence of the
$l=2$ mode $E_{\rm gw}$ on the location of the observer $R_{\rm
  obs}$. We set the observer as $R_{\rm obs} > 5000 M$ since the
inappropriate  initial condition disturbs the burst of the waveform.
From Fig.~\ref{dep-freefall-observer}, we find that the total radiated
energy for $\sigma=1.0M$ and $0.5M$ are quite different from that
found by \citet{DRPP71} and this is irrespective of the location of
the observer.  The waveform with a large standard deviation parameter
does not represent the quasi-normal ringing from a point particle.

From the comparison of our two numerical results with the two cases, we
confirm that our code has an ability to handle gravitational waves
from a particle in a free fall orbit and in a bounded orbit.  The
convergence order of our code is first order in $\sigma$ and second
order in $\Delta r_*$.  We also find that we should set the parameter
sets as $R_{\rm obs} \ge 5000M$ and $\sigma \le 0.25M$.

\section{Gravitational wave from a dust disk}
\label{sec:IV}

\subsection{Construction of a dust disk}
\label{sec:IV-1}

\begin{center}
\begin{figure}[htbp]
\includegraphics[width=0.45\textwidth]{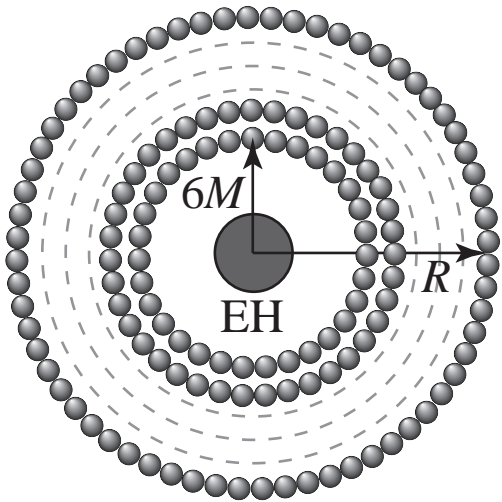}
\caption{
Illustration of our disk composed of test particles in a Schwarzschild
black hole.  EH and $R$ represent the event horizon and the outermost 
radius of the disk, respectively.
}
\label{fig_disk_image}
\end{figure}
\end{center}

Here we explain our method for constructing a dust disk in
Schwarzschild spacetime.  We make five assumptions for constructing a
dust disk as follows:
\begin{enumerate}
\item 
A disk has no self-interaction with each component; namely, a disk is
composed of test particles.
\label{itm:particle}
\item
A disk is thin so that the only components are located in the
equatorial plane of a Schwarzschild black hole.
\item
The density distribution for each radius of the disk is uniform so
that it is considered as a uniform ring for each radius.
\label{itm:ring}
\item
Each radius of the ring has a Keplerian orbit, that is to say, a
circular orbit.
\item
A disk is dynamically and radially stable.
\label{itm:stable}
\end{enumerate}
From Assumption~\ref{itm:stable}, our disk is only located
outwards of the innermost stable circular orbit of a test particle in
Schwarzschild spacetime.  We show a brief picture of our disk in
Fig.~\ref{fig_disk_image}.  Note that the disk is composed of test
particles (from Assumption~\ref{itm:particle}) in a circular orbit at
each radius.  Since we have a symmetry in the azimuthal direction of
the particle motion, we can construct a ring from test particles using
Assumption~\ref{itm:ring}.  Then, we compose a disk from rings.  

We regard the mass distribution per unit width of the disk at radius
$r$ as $m(r)$ given by
\begin{equation}
m(r) = 2 \pi C \mu r^{1-n},
\label{eqn:mdist}
\end{equation}
where $C$, $\mu$, and $n$ represent the normalization constant that
has a dimension of $r^{n-2}$, the rest mass of a test particle, and
the factor of the mass distribution, respectively.  Note that $n=0$
denotes the constant mass density distribution of the disk.  The total
mass of the disk $\mu_{\rm disk}$ is
\begin{eqnarray}
\mu_{\rm disk} &=& \int_{6M}^{R} dr ~ m(r)
\nonumber \\
&=&
\left\{ 
\begin{array}{ll}
\frac{2}{2-n} \pi C \mu [R^{2-n} - (6 M)^{2-n}] & (n \not= 2).\\
2 \pi C \mu [\ln ( R / M ) - \ln 6] & (n = 2).
\end{array}
\right.
\end{eqnarray}

In order to construct a disk numerically from a ring, we set each ring
at the location which satisfies the energy 
\begin{equation}
\tilde{E}_{\rm p}^{j} = 
\tilde{E}_{\rm p}^{1} + \frac{j-1}{N_{\rm d} -1} 
(\tilde{E}_{\rm p}^{N_{\rm d}} - \tilde{E}_{\rm p}^{1})
\quad
(j=1, \cdots, N_{\rm d})
,
\end{equation}
where $\tilde{E}_{\rm p}^1$ is the energy at $r=6M$, $\tilde{E}_{\rm
  p}^{N_{\rm d}}$ is the energy at $r=R$, $R$ is the radius of the disk,
and $N_{\rm d}$ is the number of the rings.

In order to initiate the radial motion of a dust disk, we slightly
deplete the angular momentum of each ring.  Since we maintain the
energy conserved during the depletion, we still follow the geodesic
equation for a ring after the depletion.

\subsection{Gravitational wave from a dust disk}
\label{sec:IV-2}

\begin{center}
\begin{figure}[htbp]
\includegraphics[width=0.5\textwidth]{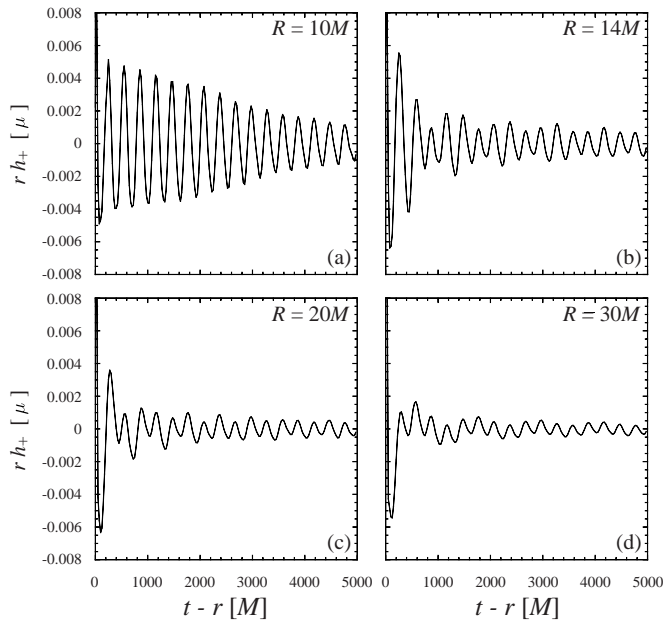}
\caption{
Gravitational waveform from a dust disk at $R_{\rm obs} = 5000M$ in
the equatorial plane with constant density for four different disk
radii, i.e., $R=$ (a) $10M$, (b) $14M$, (c) $20M$, and (d) $30M$.
}
\label{fig_n0-wave}
\end{figure}
\end{center}

Since we neglect the interaction between each component of the disk,
gravitational waves emitted from the disk are obtained by superposing
the waves emitted from each ring.  For each gravitational wave emitted
from each ring, we impose the same initial condition
(Eq.~[\ref{eqn:initial}]) and the same boundary condition
(Eqs.~[\ref{eqn:bd1}] and [\ref{eqn:bd2}]) as we used for the case of
a test particle.  In order to avoid the wave reflection at the
boundary after a long evolution, we take the spatial grid as $-5000M
\le r_* \le 10000M$.  We also set the observer at $R_{\rm
  obs}=5000M$.  The radial wave function of gravitational waves
$\Psi_{l}^{\rm (disk)}$ is given by
\begin{equation}
\Psi_{l}^{\rm (disk)} (t, r) = \frac{1}{\mu_{\rm disk}}
\int _{6M}^{R} dr_0 ~ m(r_{0}) 
\left. \Psi_{l0}^{\rm (odd/even)} \right|_{r(0)=r_{0}},
\label{emitted-GW}
\end{equation}
where $r_{0}$ represents the radius of the ring at $t=0$.  Only the
$m=0$ mode contributes to the gravitational waves from the ring due to
its axisymmetric nature, where $m$ is an index of spherical harmonics
$Y_{lm}$.  Therefore the amplitude of the gravitational wave is zero
until the angular momentum is depleted.  We only take the $l=2$ mode
into consideration because it is the dominant mode for the observer in
the equatorial plane.  In this paper we consider only radial mode due
to $m=0$. However the radial mode has a fundamental physics and there
is also in a realistic situation.  Thus it is considered that we can
extract some important evidence by this analysis.

\begin{center}
\begin{figure*}[htbp]
\includegraphics[width=0.7\textwidth]{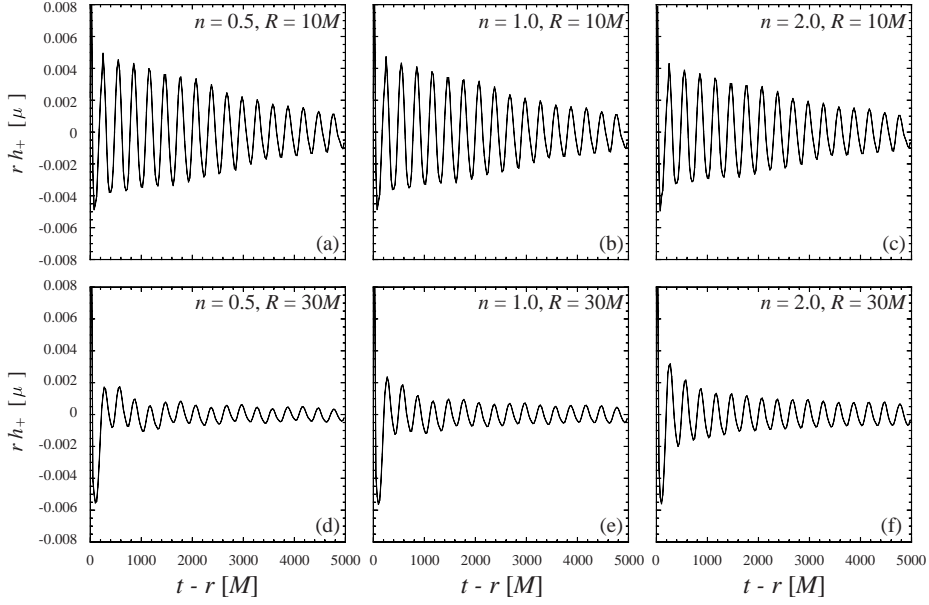}
\caption{
Gravitational waveform of a dust disk as a function of $t-r$.  The
parameters ($n$, $R$) are (a) ($0.5$, $10M$), (b) ($1.0$, $10 M$), (c)
($2.0$, $10M$),(d) ($0.5$, $30M$),(e)($1.0$, $30M$), (f) ($2.0$,
$30M$).  We set the observer in the equatorial plane at $R_{\rm obs} =
5000 M$.
}
\label{fig_wave-n-dep}
\end{figure*}
\end{center}

We define the power spectrum of the gravitational waveform as 
\begin{equation}
P(\omega) = 
\lim_{T \rightarrow \infty} \frac{1}{T} 
\left| 
  \int_{-T/2}^{T/2} [h_{+}(t) - i h_{\times}(t)] e^{- i \omega t} dt 
\right|^{2}
,
\end{equation}
where $\omega$ is a frequency.  In our computation, we choose as large
a $T$ as possible, namely $T = 4000 M$, assuming that the phase
difference apart from the periodic waves at the edge of the
integration should be negligible.  Note that we do not even integrate
one long period of the wave in Fig.~\ref{fig_n0-wave} (a) and
Fig.~\ref{fig_wave-n-dep} (a) (b) (c).  The valid frequency range of
the power spectrum is ${\rm several} \times ({2 \pi}/T) \lesssim
\omega$, namely $0.005 \lesssim M\omega$, in all our computations.
Hereafter we calculate the gravitational waves with $\Delta r_* =
\sigma = 0.25M$ and $\Delta t = 0.5 \times \Delta r_*$.

\subsubsection{Constant rest mass density distribution}
\label{sec:IV-2-1}

First we assume that the rest mass density distribution of the disk is 
uniform, namely $n=0$ in Eq.~(\ref{eqn:mdist}).  We vary the outermost
radius of the disk as $R = 10M$, $14M$, $20M$, and $30M$.  We deplete
$1\%$ of the angular momentum of each ring at $(t-r)/M=0$ to initiate
the radial motion.  As a matter of fact, the innermost radially stable
fragments of the disk approach $r \sim 7.85 M$.  We show the
gravitational waveform from a dust disk in Fig.~\ref{fig_n0-wave}.
Since a uniform ring with a circular orbit retains an axisymmetric
nature in the system, the system does not emit gravitational waves
before the depletion.

\begin{center}
\begin{figure}[htbp]
\includegraphics[width=0.5\textwidth]{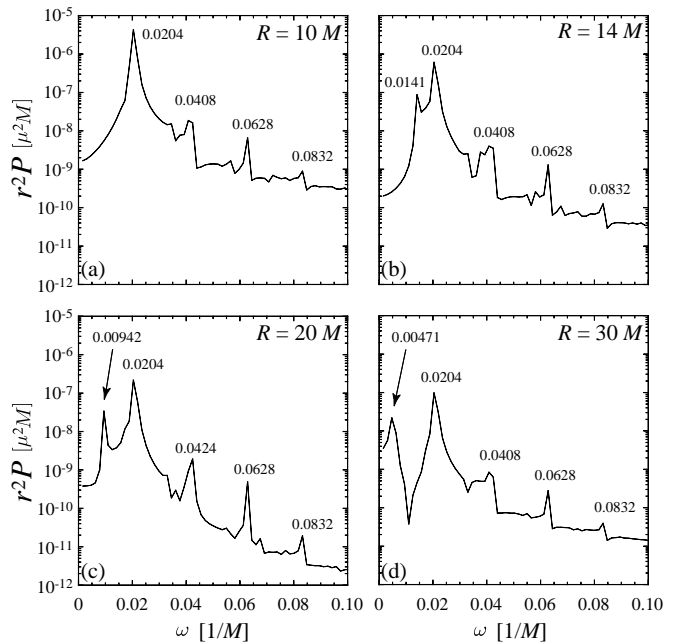}
\caption{
The power spectrum of the gravitational waveform in
Fig.~\ref{fig_n0-wave}.  We extract the time range of $1000 M \lesssim
t-r \lesssim 5000 M$ to compute the power spectrum.  The radius of the
disk $R$ is (a) $10 M$, (b) $14 M$, (c) $20 M$, (d) $30 M$.  The
values correspond to the peaks of the spectrum.
}
\label{fig_PS-n0}
\end{figure}
\end{center}

We choose $N_{\rm d} = 50$, $100$, $175$, and $300$ for $R=10M$,
$14M$, $20M$, and $30M$, respectively, so that the step-size of the
integration in Eq.~(\ref{emitted-GW}) should be the same.  As the
distance between the closest two radii of the components at $t = 0$ is
$0.035 M \sim 0.26 M$ depending on the radius, the components have
sufficient numbers of particles to compose a disk.  The first burst in
the waveform is regarded as an inappropriate choice of the initial
condition, as we mentioned in Sec.~\ref{sec:III}.

\begin{table}[htbp]
      \begin{center}
          \leavevmode
          \caption{
Relationship between the initial radius of the ring and the
characteristic frequency of gravitational waves from a ring.
          }
          \begin{tabular}{cccc}
            \hline\hline
& $R_{\rm ring}$\footnotemark[3] $[M]$ 
& $M \omega_{\rm gw}$\footnotemark[4] & \\
\hline
              & $8.0$   & $0.0204$   &  \\
              & $10.0$  & $0.0204$   &  \\
              & $14.0$  & $0.0141$   &  \\
              & $20.0$  & $0.00942$  &  \\
              & $30.0$  & $0.00471$  &  \\
            \hline\hline
          \end{tabular}
          \label{tab-omega-GW}
      \end{center}
\footnotetext[3]{$R_{\rm ring}$: Radius of the ring}
\footnotetext[4]{$M \omega_{\rm gw}$: Characteristic frequency of
  gravitational waves from a ring induced by $1\%$ depletion of the
  angular momentum.
}
\end{table}

Since the waveforms (except for the disk with $R=10M$) are very similar
to each other, it is difficult to obtain information on the disk
solely from the waveforms.  However, we find that there is a clear
difference in the power spectrum of a gravitational waveform, varying
the radius of the disk (Fig.~\ref{fig_PS-n0}).  Note that the spectrum
of Fig.~\ref{fig_PS-n0} is calculated from the waveforms of
Fig.~\ref{fig_n0-wave}.  To construct a power spectrum, we take the
data set of a waveform in the range of $1000 M \lesssim t-r \lesssim
5000 M$.  The value in Fig.~\ref{fig_PS-n0} corresponds to the
frequency $M \omega$ of each peak.  From these power spectra, the peak
frequencies are almost the same, irrespective of the disk radius.
However, for $R=14M$, $20M$, and $30M$ there is another peak
frequency, which is lower than the frequency of the first peak in the
case where $R=10M$.

\begin{center}
\begin{figure}[htbp]
\includegraphics[width=0.45\textwidth]{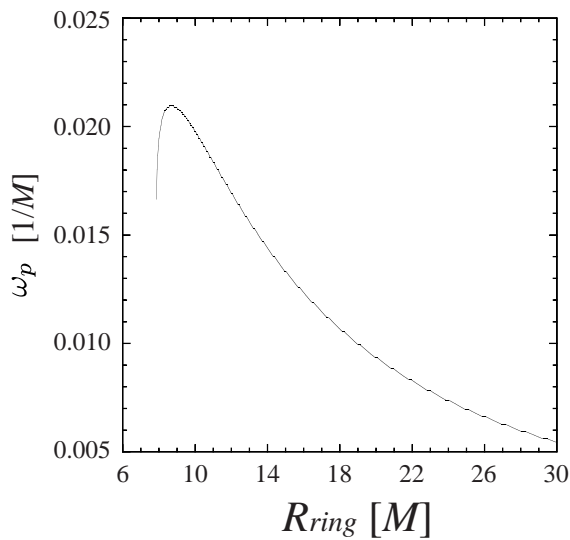}
\caption{
Radial frequency $\omega_p$ of the particle as a function of initial
location $R_{\rm ring}$.
}
\label{fig_particle}
\end{figure}
\end{center}

In order to understand the peak frequencies in the power spectra, we
calculate the characteristic frequency of a particle around a
Schwarzschild black hole.  We summarize the characteristic frequencies
of gravitational waves from a test particle with 1\% depletion of the
angular momentum from a circular orbit in Table~\ref{tab-omega-GW}.
From Fig.~\ref{fig_PS-n0} and Table~\ref{tab-omega-GW}, the peak
frequency $M \omega = 0.0204$, which appears in all disk models,
corresponds to the characteristic frequency of a ring $\omega_{\rm
  gw}$ of $R_{\rm ring} = 8M$ and $10M$, and another peak frequency
lower than $M \omega = 0.0204$ corresponds to the characteristic
frequency of a particle at the outermost edge of the disk.  Note that
the peak frequencies above $M \omega = 0.0204$ correspond to the
higher-order frequency of $M \omega = 0.0204$.  In fact, the
corresponding frequencies have the following rule: $0.0408 = 0.0204
\times 2$, $0.0628 \approx 0.0204 \times 3$, $0.0832 \approx 0.0204
\times 4$.

The peak of the frequency below $M \omega = 0.0204$ comes from the
ring at the outer edge of the disk, and the peak at $M \omega =
0.0204$ comes from the ring at the inner edge of the disk.  The reason
for no peak at the frequency below $M \omega = 0.0204$ of a disk with
$R=10M$ is that the frequency corresponding to $R=10M$ is the same as
that of the peak at $M \omega = 0.0204$.  The phase cancellation
effect plays a role in the emission of gravitational waves from the
rings between the two edges of the disk.  Therefore, it is possible to
determine the radius of a dust disk from the power spectrum if the
disk radius is $\gtrsim 12M$.

The characteristic frequency $\omega_{\rm gw}$ coincides with the
radial frequency $\omega_p$ of a particle at $r=R_{\rm ring}$ due to
the axisymmetric motion of the disk that corresponds to the $m=0$
mode.  In Fig.~\ref{fig_particle}, we plot the relation between the
radial frequencies $\omega_p$ of the particle at $R_{\rm ring}$ and
initial particle radius $R_{\rm ring}$.  Note that we deplete 1\% of
the angular momentum of the ring to initiate the radial motion.
Since our frequency resolution is $M \Delta \omega \approx 0.005$ as
we mentioned before, the frequency $\omega_p$ and $\omega_{\rm gw}$
have a good correspondence.  As mentioned above, because a peak
corresponds to $\omega_{\rm gw}$ at $R_{\rm ring}=R$ in the power
spectrum of gravitational waves, it is expected that we can determine
the radius of a dust disk using the relation between $\omega_p$ and
$R_{\rm ring}$.  The quasi-normal ringing in the waveform represents
the character of black hole; we can determine the character of the
central black hole from the ringing.  In a dust disk system, however,
we cannot find the quasi-normal ringing in the waveform because there
is a continuous inflow of fragments into a Schwarzschild black hole.
Therefore the ringing is canceled by the different phase of the wave
generated by each inflow fragment.

\begin{center}
\begin{figure*}[htbp]
\includegraphics[width=0.7\textwidth]{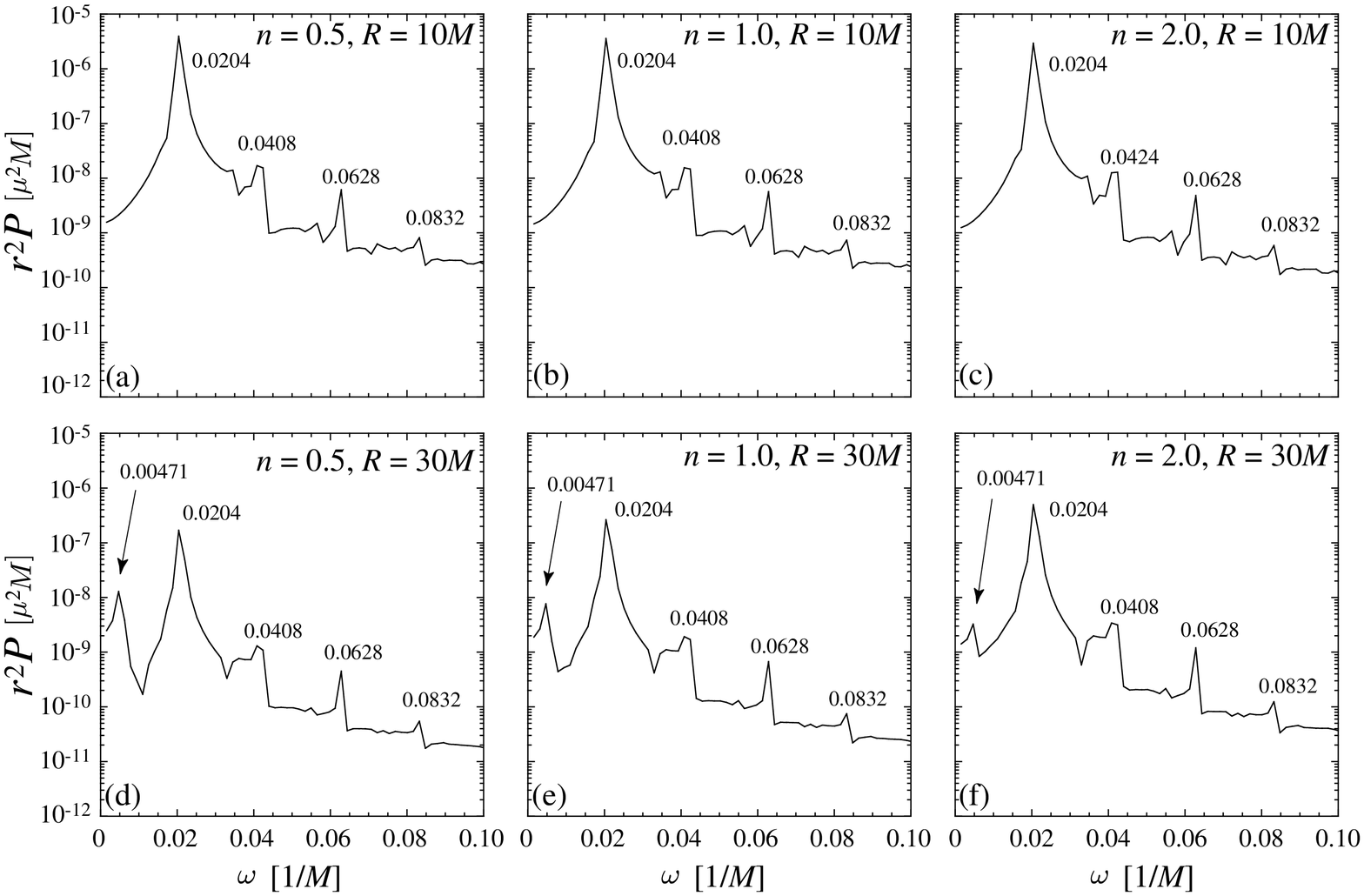}
\caption{
Power spectrum of gravitational waveforms in
Fig.~\ref{fig_wave-n-dep}.  We extract the time range of $1000 M 
\lesssim t-r \lesssim 5000 M$ to compute the power spectrum.  The
values correspond to the peaks of the spectrum. 
}
\label{fig_PS-n-dep}
\end{figure*}
\end{center}

\subsubsection{Non-constant rest mass density distribution}
\label{sec:IV-2-2}

Next we consider the non-constant mass density distribution of a dust
disk, which means that the rest mass density of the outermost edge of
the disk is lower than that of the innermost edge, i.e., $n>0$.  In
Fig.~\ref{fig_wave-n-dep}, we show a waveform from a dust disk with
$R=10M$ and $R=30M$ for $n=0.5$, $1.0$, and $2.0$, respectively.  For
a disk with $R=10M$, the waveform does not strongly depend on the
distribution.  For the disk models with $R=30M$, the waveforms
approach those of the disk models with $R=10M$ as $n$ becomes
larger.  Therefore it is also difficult to obtain information on the
disk from the observational waveforms. However, in the power 
spectra (Fig.~\ref{fig_PS-n-dep}), the frequency corresponds to the
peak having the same feature as the one in the case where $n=0$.
Thus, the peak frequencies in the power spectrum correspond to those of
gravitational waves from a ring at $R_{\rm ring}=R$ and at the inner
region of the disk.  Therefore, we conclude that we can determine the
radius of the disk irrespective of the density distribution using the
power spectrum of gravitational waves.

\begin{center}
\begin{figure}[htbp]
\includegraphics[width=0.45\textwidth]{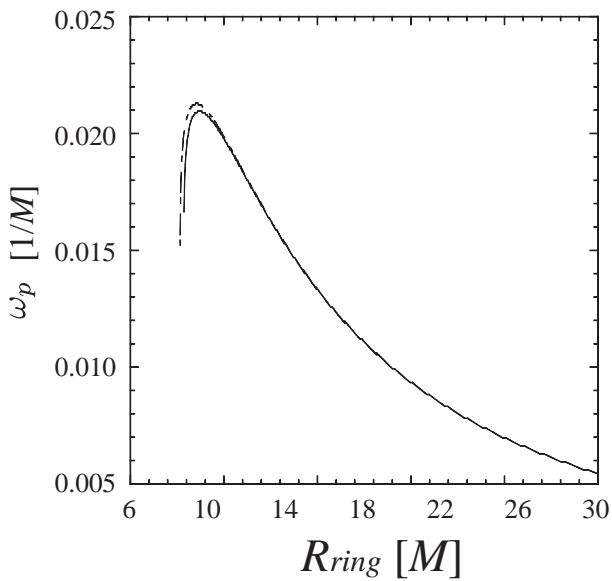}
\caption{
Radial frequency $\omega_p$ of the particle
as a function of initial location $R_{\rm ring}$.  Dash line
corresponds to $\omega_p$ of the particle with a different angular
momentum depletion (Eq.~[\ref{depletion2}]) and the solid line with
the same angular momentum depletion as Fig.~\ref{fig_particle}. 
}
\label{fig_11}
\end{figure}
\end{center}

\begin{center}
\begin{figure}[htbp]
\includegraphics[width=0.45\textwidth]{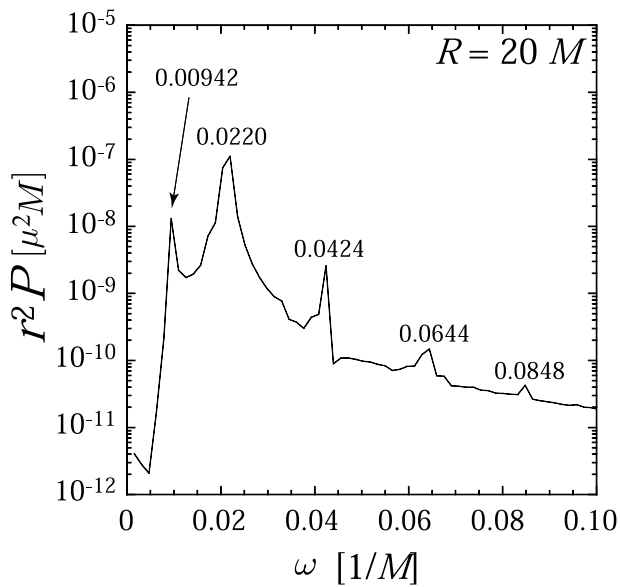}
\caption{
Power spectrum of gravitational waves from a dust disk with a
different type of angular momentum depletion (See
Eq.~[\ref{depletion2}]).  We choose the density profile of the disk as
uniform, with the disk radius $R=20M$.  The values in the figure
correspond to the peaks of spectrum. 
}
\label{fig_12}
\end{figure}
\end{center}

\subsubsection{Different type of angular momentum depletion}
\label{sec:IV-2-3}
Here we show our result based on a different type of angular momentum
depletion of test particles whether our statement largely depends on
the type of depletion.  We deplete the angular momentum of the
fragments in the following manner:
\begin{equation}
\frac{\Delta L_p}{L_p} = \frac{6M}{r} \times 10^{-2}.
\label{depletion2}
\end{equation}
The gravitational wave spectrum from the constant rest-mass density
distribution is shown in Fig.~\ref{fig_12}.  Although the orbit of
each fragments has been changed (the characteristic frequency is shown
in Fig.~\ref{fig_11}), the peak frequencies in the power spectrum
correspond to those of gravitational waves from a ring at $R_{\rm
  ring}=R$ and at the inner region of the disk.  Therefore, we can
determine the radius of the disk in a different type of angular
momentum depletion using the power spectrum of gravitational waves.

\section{Conclusions}
\label{sec:V}

We study gravitational waves from a dust disk around a Schwarzschild
black hole using a black hole perturbation approach in a time domain.
We especially focus on whether we can obtain the radius of the disk
from gravitational waves.

We find that it is difficult to obtain information on the dust disk
only by the observational waveforms.  However, it is possible to 
determine the radius of the disk using the power spectrum of
gravitational waves, irrespective of its density distribution.  Our
conclusion is quite similar to the findings of \citet{SN00,SN01}.
They claim that when a star is tidally disrupted by the central
rotating black hole, we can determine the radius of the star from the
spectrum of gravitational waves.  In addition, there is also a peak
corresponding to a frequency of gravitational waves from a particle
located at the inner edge of the disk.  Since the standard deviation
of the inner edge of the disk from $R = 6 M$ is a consequence of the
radiated angular momentum, we could estimate the amount of radiation
from the radius standard deviating from the innermost circular orbit.
Although the detail disruption process requires 3D general
relativistic calculation with the whole energy transport process, our
statement from the simple model is still helpful to understand the
disruption process by observing the gravitational wave spectrum.

We also mention the target of the gravitational wave source and its
detectability.  The typical frequency and the strength in our dust
disk model are
\begin{align}
f &= 
\frac{\omega}{2\pi} = 6.59 \times 10^{-4} 
\left( \frac{10 ^{6} M_{\odot}}{M} \right)
\left( \frac{M\omega}{0.0204} \right)
,\\
h &= 
4.78 \times 10^{-23}
\left( \frac{1 {\rm Mpc}}{r} \right)
\left( \frac{M}{10^{6} M_{\odot}} \right)
\left( \frac{\mu/M}{10^{-6}} \right)
\left( \frac{r h_{+}/\mu}{0.001} \right)
.
\end{align}
Therefore it is possible to detect gravitational waves from a dust
disk oscillating around a supermassive black hole by LISA.

We have only investigated the outermost radius of the disk up to $\sim
30M$ due to a limitation on computational time.  However, our finding
can be extrapolated into a more general astrophysical situation, such
as the black hole formation  phase of a supermassive star collapse.
In this phase, the collapse forms a supermassive black hole and a disk 
\cite{SBSS02, SS02}, and the fragments of the disk could fall
into the supermassive black hole due to some dissipative mechanism or
instabilities of the fluid that lead to a different configuration.
The interaction between each material components of the disk is
considerably small for soft equation of state such as a supermassive
star, the radial oscillation of the disk may take a dominant role in
exciting a peak in the power spectrum of gravitational waves.
Therefore it could be one source generation scenario of gravitational
waves useful for determining the size of the disk.

\begin{acknowledgments}
It is our pleasure to thank Nobuyuki Kanda and Hideyuki Tagoshi for
their valuable advice.  HS gratefully acknowledges helpful discussion
with Shijun Yoshida.  We also thank Peter Musolf for his careful
reading of our manuscript.  This work was supported in part by the
Marie Curie Incoming International Fellowships (MIF1-CT-2005-021979),
by MEXT Grant-in-Aid for young scientists (No.~200200927), and by the
PPARC grant (PPA/G/S/2002/00531) at the University of Southampton.
Numerical computations were carried out on the Pentium-4 type machines
of the Astrophysics and Cosmology Group, Waseda University, and of the
Yukawa Institute for Theoretical Physics, Kyoto University. 
\end{acknowledgments}


\end{document}